\documentclass[referee]{cjaa}           

\usepackage{graphicx}                   
\input{epsf.sty}                        
\input{psfig.sty}                       

\def\sco{Sco~X-1}
\def\1705{4U~1705--44}

\begin{document}

   \title{High Resolution and Broad Band Spectra of Low Mass
	  X-ray Binaries: A Comparison between Black Holes and Neutron Stars 
}


   \author{Tiziana Di Salvo
      \inst{1}\mailto{disalvo@fisica.unipa.it}
   \and Rosario Iaria
      \inst{1}
   \and Natale Robba
      \inst{1}
   \and Luciano Burderi
      \inst{2}
      }
   \offprints{T. Di Salvo}                   

   \institute{Dipartimento di Scienze Fisiche ed Astronomiche,
	     Universit\`a di Palermo, via Archirafi 36 - 90123 Palermo, Italy\\
             \email{disalvo@fisica.unipa.it}
        \and
             Universit\`a degli Studi di Cagliari, Dipartimento
	      di Fisica, SP Monserrato-Sestu, KM 0.7, 09042 Monserrato, Italy\\
          }

   \date{Received~~2005 month day; accepted~~2005~~month day}

   \abstract{
A common question about compact objects in high energy astrophysics is whether 
it is possible to distinguish black hole from neutron star systems with some
other property that is not the mass of the compact object. Up to now a few 
characteristics have been found which are typical of neutron stars (like quasi 
periodic oscillations at kHz frequencies or type-I X-ray bursts), but in many
respects black hole and neutron star systems show very similar behaviors.
We present here a spectral study of low mass X-ray binaries containing neutron
stars and show that these systems have spectral characteristics that are very
similar to what is found for black hole systems. This implies that it is unlikely 
we can distinguish between black holes and neutron stars from their X-ray spectra,
except for the fact that black hole systems show sometimes a more extreme 
behavior with respect to neutron star systems.
   }

   \authorrunning{T. Di Salvo et al. }            
   \titlerunning{X-ray Spectra of Low Mass X-ray Binaries }  


   \maketitle
%
%
\section{Introduction}           
\label{sect:intro}


About 10 strong X-ray sources have been established to be in binary 
systems with objects more massive than the maximum mass for a neutron 
star (NS, $\sim 3 M_\odot$, e.g. Kalogera \& Baym 1996): these are black 
hole (BH) candidates. 
Unfortunately for many X-ray binary systems the mass determination is difficult 
or impossible, because the measurement of the mass function is only possible 
when the optical counterpart is observed.  Thus some other methods have been proposed 
to identify systems containing BHs using X-ray observational properties.
The discovery of strong rapid variability of the X-ray flux in the BH X-ray
binary Cyg X--1 (Oda et al. 1971) led to the idea that such variability could
be considered a signature of the presence of a BH. 
On the basis of this idea the X-ray binary Cir X--1 was long considered 
a BH candidate. However, accreting NS can also show rapid 
variability (the radius of a NS is indeed a few Schwarzschild radii), and 
in fact Cir X--1 is now believed to contain a NS because it showed type-I X-ray 
bursts (Tennant et al. 1986).  

Other arguments are based on the X-ray spectra of accreting BHs. 
Galactic BH candidates exhibit at least two spectral states, a
`soft/high state' dominated by thermal emission at $\sim 1-2$ keV, probably 
from the accretion disk, together with a steep power law tail (with photon
index $\alpha = 2-3$), and a `hard/low state' where the accretion disk 
emission is orders of magnitude less ($0.1-0.5$ keV) and the power-law tail 
is much harder (with photon index $\alpha \sim 1.7$), so it dominates 
the energy output of the source. This hard power law shows an exponential
cutoff at temperatures of $\sim 100$~keV and is generally interpreted as
due to thermal Comptonization of soft photons in a hot electron gas close 
to the compact object.

It has been proposed that, in the framework of thermal Comptonization
models, the electron temperature of the scattering cloud ($kT_e$) should be
systematically lower for NSs than for BHs: $k T_e < 30$~keV versus 
$k T_e > 50$~keV, respectively (Tavani \& Barret 1997; Churazov et al. 1997). 
This should be a consequence of the additional cooling
provided by the NS surface, which produces soft X-ray photons and regulates
the maximum temperature achievable in these systems (Kluzniak 1993; Sunyaev \&
Titarchuk 1989). However, the discovery of some low mass X-ray binaries (LMXBs) 
containing NS with hard X-ray spectra (see e.g. Barret et al. 1991; Piraino 
et al.\ 1999), extending to energies above 200 keV has severely weakened this 
criterion, although more complicated broad band approaches have been proposed 
(e.g.\ Barret, McClintock \& Grindlay 1996, see below).

Unlike in the hard state, most of the luminosity in the soft state is due to
the soft, blackbody-like component. Recently, however, broad-band data have shown 
that a few percent of the total X-ray luminosity is emitted in the hard X-rays/soft 
$\gamma$-rays range. In this range, 
the spectra of BH X-ray binaries in the soft state can be well represented 
by a steep power-law, with photon index $\alpha \approx 2-3$, which does not have 
an observable break, at least, up to energies of the order of $m_e c^2$ 
(Phlips et al. 1996; Grove et al.\ 1998). 
Although thermal Comptonization models have been successful in describing 
the hard X-ray spectra of BH X-ray binaries in the hard states, 
they cannot easily explain the unbroken power law and the associated 
$\gamma$-ray emission observed in the soft states of BH binaries. 
For this reason the steep power-law was interpreted in terms of 
Comptonization in a converging bulk flow in the vicinity of the BH 
(e.g. Ebisawa, Titarchuk, \& Chakrabarti 1996). In fact, close to 
the event horizon, the strong gravitational field is expected to dominate 
the pressure forces, and this should give a free fall of the accreting 
material into the BH.  The inverse Comptonization of low energy photons 
from fast-moving electrons should produce the steep power law, with photon
index $\sim 2.5$ mostly determined by the mass accretion rate, observed in 
the soft states of BH binaries (Titarchuk et al. 1997; see also Titarchuk 
\& Zannias 1998). 
On the contrary, for other compact objects the pressure forces become dominant 
close to their surface, and thus a free fall state should be absent. 

According to this model, therefore, the presence of this steep power law should 
be an observational signature of a BH, because in NS systems the effect of a 
radial bulk motion should be suppressed by the radiation pressure from the NS
surface (Titarchuk \& Zannias 1998). 
However, the observation of hard power-law components in bright NS systems 
contradicts this expectation (see Di Salvo \& Stella 2002 for a review, see 
also \S~2), and suggests to prefer models which do not rely on the presence
of an event horizon in the system.

Other methods have been proposed to distinguish between NSs and BHs. 
In particular, Barret et al.  (1996) proposed a luminosity criterion.
They compared the 1-20 keV luminosity to
the 20-200 keV luminosity for all BHs (with mass function estimates 
indicating a mass of the compact object larger than $3\; M_\odot$)  
and all NSs of the atoll class (the so-called X-ray bursters) detected up 
to at least 100 keV.  They find a clear distinction in
luminosity between the X-ray bursters, which lie in the so-called {\it X-ray
burster box}, and the BHs, which are found outside.  However, the distinction 
between NSs and BHs is no more evident when the Z sources, in which a hard 
tail has been detected, are included in the diagram (see Di Salvo et al.\ 2001).  

Another important spectral characteristics of BH candidates is that they often show
the presence of broad and skewed features at the energy of the iron K-shell
fluorescence line and edge. In fact, many (both stellar mass and supermassive) 
BHs  show very broad emission lines at around 6 keV, which suggest these are
produced very close to a fast spinning (Kerr) BH (see e.g.\ MCG--6--30--15, 
Fabian et al.\ 2002; GX 339--4, Miller et al.\ 2004; XTE J1650--500, Miniutti 
et al.\ 2004). On the other hand, not much is known about iron K-shell emission
in LMXBs containing NS. These systems often show the presence of broad
iron lines. However, such studies are performed mainly with low-energy 
resolution instruments, and need to be confirmed by a high-resolution 
spectroscopy. 

We present here a spectral study of a sample of LMXBs containing NS, both
in a broad band X-ray range with INTEGRAL and RXTE, and with high energy
resolution using the HETG on board of Chandra. In particular we show in \S~3 the
presence of hard X-ray emission in \sco, which does not require a high energy
cutoff up to 200 keV strongly suggesting a non-thermal origin of this component.
We also present in \S~4 high resolution spectra of \1705, which shows an intrinsically
broad iron line probably produced in the innermost rim of the accretion disk.
In the next section, \S~2, we give an introduction on the characteristics in 
the X-ray range of LMXBs hosting an old, low magnetized, accreting NS.

\section{Neutron star low mass X-ray binary basics}
\label{sect:Obs}

The modern classification of LMXBs relies upon the branching displayed by
individual sources in the X-ray color-color diagram (CD) assembled by using the
sources' count rate over a "classical" X-ray energy range (typically 2-20 keV).
This classification has proven very successful in relating the spectral and
time variability properties (see e.g.\ Hasinger \& van der Klis 1989; for 
a review see van der Klis 2000) depending
on the pattern described by each source in the X-ray CD.
It comprises a Z-class (source luminosities close to the Eddington
luminosity, $L_{\rm Edd}$) and an atoll-class (luminosities of
$\sim 0.01-0.1\ L_{\rm Edd}$).
Most atolls emit Type-I X-ray bursts, i.e. thermonuclear flashes in the layers 
of freshly accreted material onto the NS surface; only two
Z-sources are (somewhat peculiar) X-ray bursters.
Considerable evidence has been found that the mass accretion rate
(but not necessarily the X-ray luminosity)
of individual Z-sources increases from the top left to the
bottom right of the Z-pattern ({\it e.g.} Hasinger et al. 1990),
i.e. along the so called horizontal, normal and flaring branches
(hereafter HB, NB and FB, respectively). Similarly in atoll sources
the accretion rate increases from the so-called island to the top of the
upper-banana branch.

Hard X-ray components extending up to energies of several hundred
keV have been revealed in about 20 NS LMXBs of the atoll class.
In these systems the power law-like component, with typical slopes
of $\Gamma \sim 1.5 -2.5$, is followed by an exponential cutoff,
the energy of which is often in between $\sim 20$ and many tens of keV.
This component is interpreted in terms of unsaturated thermal Comptonisation.
There are instances in which no evidence for a cutoff
is found up to $\sim 100-200$~keV. This is the so called "hard state"
of atoll sources, and is similar to the hard state of BH binaries. 
There are sources that appear to spend most of the time
in this state (e.g.\ 4U 0614+091, Piraino et al. 1999, and
references therein). In others a gradual transition from the soft to the
hard state has been observed in response to a decrease of the source
X-ray luminosity and/or the source drifting from the banana branch
to the island state. This transition is often modelled in terms of a gradual 
decrease of the electron temperature of the Comptonising region.

As first noted by van Paradijs \& van der Klis (1994), there is a
clear trend for the spectral hardness of these sources (and accreting
X-ray sources in general) over the 13-25 and 40-80~keV energy ranges to be
higher for lower X-ray luminosity. This is in agreement with the observation
that Z-sources usually show much softer X-ray spectra with characteristic 
temperatures of 3--6 keV. However, recent broad band studies, mainly performed
with BeppoSAX (0.1--200 keV) and RXTE (2--200 keV), have shown that 
many Z-sources display variable hard, power-law shaped components, dominating
their spectra above $\sim 30$~keV.
Hard emission in Z-source spectra were occasionally detected
in the past. The first detection was in the spectrum of Sco X--1; beside the
main X--ray component (equivalent bremsstrahlung temperature of $\sim 4$~keV)
Peterson \& Jacobson (1966) found a hard component dominating the spectrum
above 40~keV.
The latter component was observed to vary by as much as a factor of 3.
More recently the presence of a variable hard tail
in Sco X--1 was confirmed by OSSE and RXTE observations
(Strickman \& Barret 2000; D'Amico et al. 2001).
A hard X-ray emission was also found in the Z-sources GX 17+2 (Di Salvo et al. 2000),
GX~349+2 (Di Salvo et al. 2001) and Cyg~X-2 (Di Salvo et al. 2002), as well as the 
peculiar bright LMXB Cir X--1 (Iaria et al. 2001, 2002) and during type II
bursts from the Rapid Burster (Masetti et al. 2000).
The fact that a similar hard component has been observed in several Z sources 
indicates that this is probably a common feature of these sources.
This hard component can be fitted by a power law, with photon index in the
range 1.9--3.3, contributing up to 10\% of the source bolometric luminosity.
The presence of the hard component in Z sources is in some cases related to
the source state or its position in the CD.
This was unambiguously shown for the first time by the BeppoSAX observation 
of GX~17+2 (Di Salvo et al. 2000), where the hard tail was observed
to vary systematically with the position of the source in the CD.
In particular the hard component (a power-law with photon index of $\sim 2.7$)
showed the strongest intensity in the HB of its CD; a factor of $\sim 20$ 
decrease was observed when the source moved from the HB to the NB, i.e.\ from 
low to high inferred mass accretion rate.

However, the origin of this hard component is still poorly understood.
While in most cases the hard component becomes weaker at higher
accretion rates, HEXTE observations of Sco X--1 showed a hard power-law
tail in 5 out of 16 observations, without any clear correlation with the
position in the CD (D'Amico et al.\ 2001). Also, the thermal vs.\ non-thermal
nature of this component remains to be addressed, yielding important
information on the production mechanism.
We report in the next section on hard X-ray observations of \sco\ performed 
simultaneously with INTEGRAL and RXTE. The INTEGRAL spectrum shows with high statistical
significance the presence of a hard (power-law) component, without any
clear exponential cutoff up to $\sim 200$ keV. Contrary to what is found by 
D'Amico et al.\ (2001), the intensity of this component seems to be
correlated with the source position in the CD, showing a phenomenology that is 
similar to GX 17+2.

As already mentioned, another important common feature observed in the spectra 
of X-ray binaries, both in systems containing BHs and in systems hosting an old 
accreting NS, is given by the discrete features from the iron K-shell: broad emission  
lines (FWHM up to  $\sim 1$ keV)  at energies in the range 6.4 -- 6.7 keV have been
often observed. These lines are identified with  the K$\alpha$ radiative 
transitions of iron at different  ionization stages. Sometimes
an iron  absorption edge at energies  $\sim 7-8$ keV  has been detected.
These features are powerful tools to investigate the structure
of the accretion flow  close to  the central  source; in particular,
important information can be obtained from detailed
spectroscopy of the  iron K$\alpha$ emission line and absorption edge,
since  these  are determined by the ionization stage, geometry, and
velocity field of the reprocessing  plasma.

To explain the large width of these  lines it has been proposed that  they
originate from emission reprocessed by the accretion disc surface
illuminated by the  primary  Comptonized spectrum (Fabian et al. 1989).
In  this  model, the combination of relativistic Doppler effects arising
from the high orbital velocities  and gravitational effects  due to  the
strong field in  the vicinity of the NS smears
the  reflected features. Therefore the  line will have a characteristically
broad profile, the detailed shape of which depends on the inclination
and  on  how deep  the  accretion disk  extends  into  the NS
potential (e.g.\ Fabian et al. 1989). In the case of some
BH systems, these features are so broad that they require an almost 
maximally spinning (Kerr) BH.

An alternative location of the line emitting region is the inner part
of the so-called Accretion Disk Corona (ADC), probably formed by
evaporation of the outer layers of the disk illuminated by the emission
of the central object (e.g.\ White \& Holt 1982).
In this case the width of the line is explained by
thermal Comptonization of the line photons in the ADC. This produces
a genuinely broad Gaussian distribution of line photons, with
$\sigma \geq E_{\it Fe} (k T_e/m_e c^2)^{1/2}$,
where $E_{\it Fe}$ is the centroid energy of the iron line and $k T_e$ is
the electron temperature in the ADC (see Kallman \& White 1989;
Brandt \& Matt 1994 for more detailed calculations). This mechanism can
explain the width of the iron line for temperatures of the emitting
region of few keV.
The presence of several unresolved components, which can eventually be
resolved by the high resolution X-ray instruments on board Chandra and
XMM-Newton, can also contribute to broaden the line.

In \S~\ref{sect:1705} we present the results of a Chandra observation of
the atoll source \1705; we selected this source for a Chandra observation because 
a broad (1.1 keV FWHM) iron emission line at 6.5 keV has been previously reported
(White et al. 1986; Barret \& Olive 2002). One of the goals of our Chandra 
observation was to study the iron line profile to discriminate among the various 
models that have been proposed to explain the large line width. The Chandra/HETGS
observation demonstrates that the iron line is intrinsically broad
(1.2 keV FWHM).

\section{Scorpius X-1 with INTEGRAL: non-Thermal Hard Emission?}
\label{sect:data}

To study the hard X-ray emission in the brightest of these sources, as well as
its correlation with other source properties (such as radio emission
and fast timing variability) we have performed a campaign of
observations of Sco X--1 with INTEGRAL and RXTE. Part of these observations
were also performed simultaneously with radio VLBI observations (which will
be discussed elsewhere).
Sco X-1 was observed during two complete INTEGRAL revolutions on 2003 July 30
-- August 1 and 2003 August 11 -- 13, simultaneously with RXTE (Di Salvo et al.,
in preparation). The X-ray CD of Sco X-1 during the INTEGRAL observations is
shown in Figure~1 (left panel). During each of the two observations the source 
described a fairly complete Z-track in the CD.
%
\begin{figure}
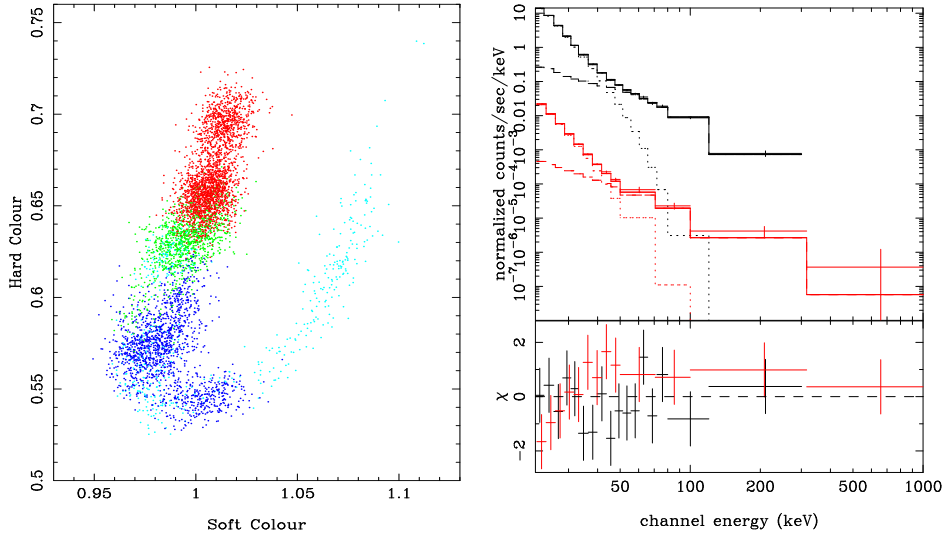

   \vspace{2mm}
   \begin{center}
   \hspace{3mm}\psfig{figure=disalvo_2005_01_fig1a.ps,width=60mm,height=70mm,angle=0.0}
   \hspace{3mm}\psfig{figure=disalvo_2005_01_fig1b.ps,width=60mm,height=70mm,angle=0.0}
   \caption{ {\bf Left:} Color-color diagram of Sco X--1 from PCA data during the
simultaneous INTEGRAL/RXTE observation. Here Soft Color is the ratio
of the count rate in the energy bands [$3.5-6$ keV]/[$2-3.5$ keV] and
Hard Color is the ratio [$9.7-16$ keV]/[$6-9.7$ keV], respectively.
Only data form PCUs 2 and 3, which were on for most of the observation,
have been used. The four colors indicate the four regions in which the CD has
been divided and from which INTEGRAL spectra were extracted. 
{\bf Right:} Top: ISGRI (20 -- 200 keV, black) and SPI (20 keV -- 1 MeV,
red) data of \sco\ together with the best-fit model (solid lines on top
of the data) composed of the Comptonization model ({\tt comptt}, dotted
lines) and a cutoff power law (dashed lines).
Bottom: residuals in units of $\sigma$ with respect to the best fit model.   }
   \label{Fig:scox1}
   \end{center}
\end{figure}

In order to check for the presence of the hard tail in our data, we first 
analysed the INTEGRAL (IBIS/ISGRI and SPI, energy band 20 keV -- 1 MeV) 
spectrum integrated over the whole observation.
Similar to what has been done for other LMXBs of the Z-class, we fitted
the INTEGRAL spectrum with the Comptonization model {\tt compTT}
(Titarchuk 1994).
This model gives a good fit of the soft part of the \sco\ spectrum up to
$\sim 40-50$ keV. Above this energy, a hard excess is clearly visible in
the residuals independent of the particular Comptonization model used to fit
the soft part of the spectrum. The fit is significantly improved by adding
to the {\tt compTT} model a power-law with photon index $\sim 2.7$
(this gives a reduction of the $\chi^2$/dof from 892/27 to 42/25).
We tested the presence of a thermal cutoff in the hard power-law;
substituting the power law with a cutoff power-law does not improve the
fit significantly (the latter model gives a $\chi^2$/dof = 41/24), and
the temperature of the exponential cutoff is $k T > 200$ keV ($90\%$
confidence level).
The best-fit parameters for the ISGRI (20 -- 300 keV) and SPI (20 keV --
1 MeV) spectra are reported in Table~1; data and residuals with respect
to the best fit model are shown in Figure~1 (right panel).
\begin{table}
\footnotesize
\caption{Results of the fitting of the \sco\ INTEGRAL/ISGRI (20--200 keV)
spectra resolved along the Z-track in the CD. The total spectrum includes
SPI data and is fitted in the 20 keV -- 1 MeV range.}
\label{table:1}
\newcommand{\m}{\hphantom{$-$}}
\newcommand{\cc}[1]{\multicolumn{1}{c}{#1}}
\renewcommand{\tabcolsep}{0.6pc} 
\renewcommand{\arraystretch}{1.2} 
\begin{center}
\begin{tabular}{@{}llllll}
\hline
Parameter               & HB/UNB & NB & NB/FB & FB & Total \\
\hline
$k T_0$ (keV)                 & $1.3$ (frozen) & $1.3$ (frozen) &
                                $1.3$ (frozen) & $1.3$ (frozen) & $1.3$ (frozen) \\
$k T_{\rm e}$ (keV)           & $3.34 \pm 0.03$ & $3.58^{+0.17}_{-0.06}$ &
                                $3.82^{+0.06}_{-0.10}$ & $3.4^{+0.5}_{-0.2}$ &
				$4.73^{+0.03}_{-0.15}$\\
$\tau$                        & $5.19^{+0.18}_{-0.28}$ & $4.40^{+0.58}_{-0.15}$ &
                                $3.73 \pm 0.15$ & $4.7 \pm 1.0$ & $2.43 \pm 0.04$ \\
PhoIndex                      & $2.73^{+0.06}_{-0.12}$ & $2.59^{+0.30}_{-0.76}$ &
                                $2.86^{+0.07}_{-0.24}$ & $2.7$ (frozen) & 
				$2.30^{+0.01}_{-0.23}$ \\
$kT$ (keV)                    & $> 290$ & $> 94$ & $> 140$ & -- & $> 226$ \\
Flux (20--40)   & $7.26 \pm 0.22$  & $5.86 \pm 0.37$ & $4.36 \pm 0.50$  & 
		$7.07 \pm 0.45$ & $5.86 \pm 0.16$ \\
Flux (40--200)  & $4.8 \pm 1.6$ & $3.0 \pm 1.8$ & $2.19 \pm 0.89$ & 
		$0.57 \pm 0.36$ & $3.3 \pm 1.1$ \\
$\chi^2(d.o.f.)$                            & $9.4 / 11$ & $15.0 / 11$
                                    & $16.1 / 11$ & $9.85 / 13$ &  $24.4 / 23$ \\
\hline
\end{tabular}\\[2pt]
\end{center}
\footnotesize
The model consists of a Comptonized spectrum modeled by {\tt comptt}, and
a cutoff power law.
$k T_0$ is the temperature of the seed photon (Wien) spectrum, $k T_e$ the
electron temperature and $\tau$ the optical depth in a spherical geometry.
Fluxes are given in units of $10^{-9}$ erg cm$^{-2}$ s$^{-1}$ in the range
20 -- 40 keV and in units of $10^{-10}$ erg cm$^{-2}$ s$^{-1}$ in the range
40 -- 200 keV.
All the uncertainties are calculated at 90\% confidence level.
\end{table}

To look for variability in the hard component with the spectral state
of the source, as measured by its position in the X-ray CD, we divided
the Z-track in the CD of \sco\ into four parts corresponding to
the HB/upper-NB, the NB, the NB/FB vertex, and the FB, respectively. 
We therefore extracted INTEGRAL/ISGRI spectra for each of the time 
intervals mentioned above, resulting in four CD resolved spectra.
Unfortunately there was no superposition between RXTE and INTEGRAL data
in the FB during the first part of the observation).
The INTEGRAL/ISGRI exposure times for the four intervals were 52.4 ks,
45.5 ks, 60.6 ks and 6 ks.
We fitted each of this spectra with {\tt comptt} and a cutoff power law.
This model gave a good fit of the first three spectra, with little variability
in the spectral parameters (see Table~1).
For each of these spectra, the hard power law component was required in
order to fit the data.
Remarkably the FB spectrum did not require a power law
component and could be fitted with a simple Comptonization model.
Including the power law in the spectral fit of the FB
spectrum, with the photon index fixed at 2.7, a good fit requires a
decrease of the normalization of the power law component by a factor
at least 5 with respect to the average spectrum.

\section{A broad iron line in the Chandra/HETG spectrum of \1705}
\label{sect:1705}

\1705 was observed using the High-Energy Transmission Grating Spectrometer
(HETGS) on board of Chandra starting on 2001 July 1 (see Di Salvo et al.\ 2005
for details).
We fit the HEG first order spectra of \1705 to a continuum model. The best
fit model consists of the Comptonization model {\tt comptt}
(Titarchuk 1994), modified by absorption from neutral matter, parametrized
by the equivalent hydrogen column $N_H$, which gives a
$\chi^2_{\rm red} (d.o.f.)$ of $1.05 (3265)$. This model also includes
an overabundance of Si by a factor $\sim 2$ with respect to Solar
abundances to fit a highly significant absorption edge at
$\sim 1.84$ keV (the addition of this parameter reduces the $\chi^2$ by
$\Delta \chi^2 \simeq 61$ at the expense of 1 degree of freedom).
Note that we cannot exclude that this feature may be due to the presence of
Si in the CCDs, and therefore this overabundance is not discussed further.
Finally, in all the fits we include an
instrumental feature at $2.06$ keV (usually present in the HETG spectra
of bright sources, see Miller et al. 2002) which is fitted by an inverse
edge (with $\tau \sim -0.1$).

Residuals in units of $\sigma$ with respect to the continuum model
described above are shown in Figure~2 (right panel); several discrete features are
still clearly visible in the residuals with respect to this continuum
model, at $\sim 1.5$, $2.0$, $2.6$ keV, and, particularly, in the 6--7 keV
range, where the K$\alpha$ iron emission line is expected.
From these residuals it is apparent that the iron line is intrinsically broad
and shows a complex profile.

%
\begin{figure}
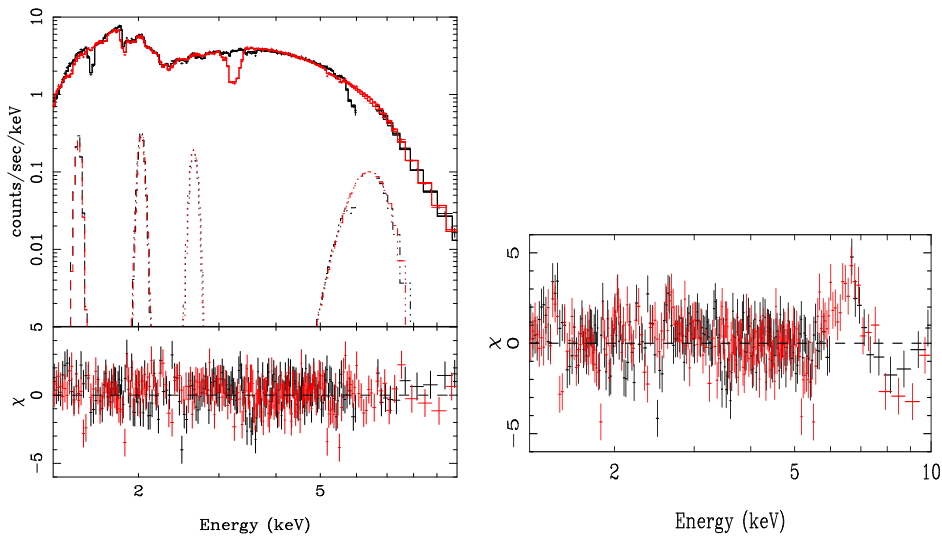

   \vspace{2mm}
   \begin{center}
   \hspace{3mm}\psfig{figure=disalvo_2005_01_fig2a.ps,width=60mm,height=70mm,angle=0.0}
   \hspace{3mm}\psfig{figure=disalvo_2005_01_fig2b.ps,width=60mm,height=40mm,angle=0.0}
   \caption{ {\bf Left:} Top panel: HEG first order spectra of \1705 together with the
best-fit model (see Table 2). The discrete features included in the
best-fit model and described by Gaussians are also shown. 
Bottom panel: Residuals in units of $\sigma$ with respect to the best fit
model.
{\bf Right:} Residuals in units of $\sigma$ with respect to the simple
Comptonization continuum when the iron line and other low energy Gaussians
are not included in the model.   }
   \label{Fig:1705}
   \end{center}
\end{figure}

The addition of a broad ($\sigma \sim 0.5$ keV) Gaussian line centered
at 6.5 keV proves necessary, giving $\Delta \chi^2 = 167$ for the
addition of three parameters. We also added three narrow emission lines
to fit the other low energy residuals mentioned above.
The addition of Gaussian emission lines at $\sim 1.5$, $2.0$, $2.6$ keV
gives a reduction of the $\chi^2$ by 27, 29, and 32 units, respectively,
for the addition of three parameters. The errors in the normalizations of these
features give a detection at about $3 \sigma$ confidence level. This is not a
highly significant detection, and needs a confirmation with future
observations. However, the fact that the energies of these features are
close to the energies of Ly$\alpha$ transitions of H-like ionization stages
of the most abundant ions emitting in the observed range (that are Mg XII,
Si XIV, and S XVI, respectively) adds further confidence that these lines may
be real.
Data and residuals in units of $\sigma$ with respect to this best-fit model
are shown in Figure~2 (left panel), and the best fit model is reported in Table 2.

In analogy with what is done for BH spectra, we also try to fit the broad 
emission line at $\sim 6.5$ keV with the line profile expected from a thin Keplerian
accretion disk. Substituting the Gaussian line with the {\tt diskline}
model, we obtain an equivalently good fit.
The line best fit parameters for the {\tt diskline} model are given in Table~3.
Therefore, the iron line observed in \1705\ is compatible with a relativistic line
produced by reflection in a cold accretion disk. In this case, we
estimate that the required inner radius of the disk is $\sim 7\; R_g$
or $\sim 15$ km for a $1.4\; M_\odot$ NS. Note that the quite
small inner radius of the disk inferred from this model is in agreement
with the quite soft X-ray spectrum of \1705 during the Chandra observation,
which would probably place the source in the banana state of its X-ray
CD. In this model, the inclination of the disk with respect to the line of
sight is constrained in the range $55^\circ - 84^\circ$.

\begin{table}[h!]
\footnotesize
\caption{Results of the fitting of the \1705 HEG first order spectra in the
1.3--10~keV energy band.  }
\label{table:1}
\newcommand{\m}{\hphantom{$-$}}
\newcommand{\cc}[1]{\multicolumn{1}{c}{#1}}
\renewcommand{\tabcolsep}{1.5pc} 
\renewcommand{\arraystretch}{1.2} 
\begin{center}
\begin{tabular}{@{}ll}
\hline
Parameter               & Value \\
\hline
$N_{\rm H}$ $\rm (\times 10^{22}\;cm^{-2})$ & $1.42 \pm 0.06$ \\
Si / Si$_\odot$                             & $2.0 \pm 0.2$ \\
$k T_0$ (keV)                               & $0.50 \pm 0.02$ \\
$k T_{\rm e}$ (keV)                         & $2.29 \pm 0.09$  \\
$\tau$                                      & $17.7 \pm 0.7$  \\
$E_{\rm Fe}$ (keV)                          & $6.54 \pm 0.07$ \\
$\sigma_{\rm Fe}$ (keV)                     & $0.51 \pm 0.08$ \\
I$_{\rm Fe}$ ($10^{-2}$ cm$^{-2}$ s$^{-1}$) & $1.5 \pm 0.3$  \\
EW$_{\rm Fe}$ (eV)                          & $ 170 $ \\
Flux (1.3--10 keV, erg cm$^{-2}$ s$^{-1}$)  & $7.82 \times 10^{-9}$ \\
Final $\chi^2(d.o.f.)$                      & $3168 / 3255$ \\
\hline
\end{tabular}\\[2pt]
\end{center}
The model consists of a Comptonized spectrum modeled by {\tt comptt}, and
four Gaussian emission lines.
$k T_0$ is the temperature of the seed photon (Wien) spectrum, $k T_e$ the
electron temperature and $\tau$ the optical depth in a spherical geometry.
For the discrete features, $I$ is the intensity of the line and $EW$ is
the corresponding equivalent width. For the parameters of the other discrete 
features included in the fit see Di Salvo et al.\ (2005). 
Uncertainties are given at 90\% confidence level.
\end{table}


\begin{table}[htb]
\caption{Iron line parameters from the {\tt diskline} model.}
\label{table:2}
\newcommand{\m}{\hphantom{$-$}}
\newcommand{\cc}[1]{\multicolumn{1}{c}{#1}}
\renewcommand{\tabcolsep}{2pc} 
\renewcommand{\arraystretch}{1.2} 
\begin{center}
\begin{tabular}{@{}ll}
\hline
Parameter               & Value \\
\hline
Energy (keV)            & $6.40 \pm 0.04$ \\
R$_{\rm in}$ (R$_g$)    & $7^{+4}_{-1}$ $(< 11)$ \\
R$_{\rm out}$ (R$_g$)   & $410^{+230}_{-130}$ \\
Inclination (deg)       & $59^{+25}_{-4}$ \\
Index                   & $2.1 \pm 0.2$ \\
I ($10^{-2}$ cm$^{-2}$ s$^{-1}$) & $1.8 \pm 0.3$ \\
Final $\chi^2(d.o.f.)$  & $3164/3252$ \\
\hline
\end{tabular}\\[2pt]
\end{center}
The other best fit parameters are compatible with those reported in Table 1.
Index refers to the power-law dependence of emissivity which scales
as $r^{-\rm Index}$.
Uncertainties are 90\% confidence level for a single parameter of interest.
\end{table}

\section{Conclusions}
\label{sect:conclusion}

In this paper we present strong evidences of tight analogies between BH and
NS X-ray spectra. In analogy with the steep and hard power-law components 
detected in some soft states (i.e.\ intermediate and very high) of BH X-ray
binaries, we show here that most Z sources, which are always found in a soft 
state, show similar components. 
In particular, the INTEGRAL observation of \sco\ shows that a steep power
law (with photon index 2.7) dominates the source spectrum above 
$\sim 30$ keV, which does not show evidence of a high energy cutoff up to
200 keV. The presence of such a hot plasma in a system which emits most of the
energy as soft X-ray photons is unlikely, because of the strong Compton cooling
expected. Therefore we prefer the interpretation of the hard power law as
non-thermal Comptonization, probably on fast moving electrons that are part of
an outflow or a jet. We have also presented the first detection in a NS LMXB,
\1705, of an intrinsically broad emission feature at the energy of the iron
K-shell fluorescence line, very similar to the broad iron lines detected in
some (stellar-mass and supermassive) BHs. We show that the feature observed 
by Chandra in \1705\ is probably originating by reflection of the primary 
spectrum in a cold accretion disk. In this case, the inner radius of the disk 
required to explain the width of the line is very close to the NS surface 
(about 1.5 NS radii). 
In this sense, the behavior of BH candidates can be different from (more extreme than) 
the behavior of NS X-ray binaries, since the iron line observed in BHs are often
so broad that they require an almost maximally spinning BH.

\begin{acknowledgements}
This work was partially supported by the Ministero della Istruzione,
della Universit\`a e della Ricerca (MIUR). TD wants to thank all the
people who have contributed to find the results presented here, and 
in particular P. Goldoni, P. Sizun, L. Stella, and M. van der Klis.  
\end{acknowledgements}

\bigskip
\noindent
{\b DISCUSSION}
\bigskip

\noindent
{\b WOLFGANG KUNDT:} When dealing with black hole (BH) candidates you tended to 
speak of "black holes" (rather than neutron stars surrounded by massive disks).
In my 2004 Springer book I argue that BH-model is multiply inconsistent. 
\bigskip

\noindent
{DI SALVO TIZIANA:} Yes, you are right, I just used the model that is most 
commonly accepted at the moment.
\bigskip

\noindent
{\b FRANCO GIOVANNELLI:} In which way you determine the angle of inclination 
of the accretion disk around the neutron star? and do you suppose that the
accretion disk plane is co-planar to the orbital plane?
\bigskip

\noindent
{DI SALVO TIZIANA:} The inclination angle is determined for \1705 from
the shape of the iron line modelled by a diskline; indeed the width and
the distortion of the line depends on the Keplerian velocy along the line 
of sight, and therefore on the inclination of the disk with respect to the 
observer. The hypothesis that accretion disk is co-planar to the orbital plane
is not necessary in this case.
\bigskip

\label{lastpage}

\end{document}